\documentstyle[twoside,fleqn,espcrc2,epsfig]{article}

\newcommand{\AmS}{{\protect\the\textfont2
  A\kern-.1667em\lower.5ex\hbox{M}\kern-.125emS}}

\title{The anisotropy of inverse beta decay and antineutrino detection}
\author{G.Fiorentini, M.Moretti, F.L.Villante
\\
\smallskip
\smallskip
Dipartimento di Fisica, Universita' di Ferrara, I-44100, Ferrara
 (Italy)\\
Istituto Nazionale di Fisica Nucleare, Sez. di Ferrara, I-44100,
 Ferrara
 (Italy)}

\begin{document}

\begin{abstract}
{The anisotropy of 
 the positrons emitted in the reaction
$\overline{\nu}_{e}+p\rightarrow n+e^{+}$ has
to be taken into account for extracting an antineutrino signal
in Superkamiokande. For the Sun, this effect
allows a sensitivity to  
 $\nu_{e}\rightarrow\overline{\nu}_{e}$
 transition probability at the 3\%
level already with the statistics collected in 
the first hundred days.
 For a supernova in the Galaxy, the effect
is crucial for extracting the correct ratio of 
$\nu-e$ to $\overline\nu_{e}-p$ events.}
\end{abstract}
\maketitle

As well known, the specific signature of antineutrinos in hydrogen
containing materials is through the inverse beta decay (I$\beta$D),
$\overline{\nu}_{e}+p\rightarrow n+e^{+}$,
which produces {\it almost isotropically} distributed monoenergetic 
positrons ($E_{e^{+}}=E_{\overline{\nu}}-\Delta m; \Delta m=m_{n}-m_{p}$).
For energy above a few MeV, the differential cross section
is:

\begin{equation}
\frac{d\sigma}{d\cos\theta}=\frac{\sigma_0(E_{\overline{\nu}})}{2}
(1-a\cos\theta)
\label {angdep}
\end{equation}

%\noindent
%where:
%
%\begin{equation}
%\sigma_{0}(E_{\overline{\nu}})=9.2\times10^{-42} cm^{2}
%[(E_{\overline{\nu}}-\Delta m)/10 MeV]^{2}
%\end {equation}
%
where:
\begin{equation}
 a=\frac{(g_{A}/g_{V})^{2}-1}{3(g_{A}/g_{V})^{2}+1}\simeq0.1
\quad ,
\end{equation}
$g_{V}$ ($g_{A}$) is the vector (axial)
coupling of the neutron and $\sigma_0(E_{\overline{\nu}})$
is the total cross section for antineutrino
energy $E_{\overline{\nu}}$.

The goal of this paper is to assess the  importance
of the angular dependence of the cross section in two situations
of practical interest:
\begin{itemize}
\item The search for solar antineutrinos;\footnote{This
signal is of interest since it might be a signature
of $\nu$ decay or magnetic moment. {\em see \cite{fvl} for details and
references}}
\item the detection of a supernova in the Galaxy.
\end{itemize}

\section {Solar Antineutrinos}

\begin{figure}[htb]
\begin{center}
\vspace{-1.0cm}
\epsfig{file=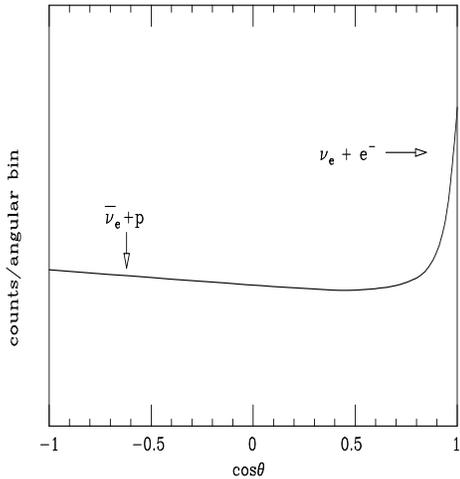,height=8cm,width=8cm,angle=90}
\vspace{-1.7cm}
\end{center}
\caption[aaa]{Sketch of the expected angular distribution of events in
the presence of a solar $\overline{\nu}_{e}$ flux.}
\label{fig1}
\end{figure}

In the absence of a solar antineutrino flux, the Superkamiokande (SK)
 background is expected
to be isotropic. In the presence of solar antineutrinos,
positrons emitted by I$\beta$D contribute to the background,
which, due to the angular dependence of I$\beta$D cross section,
 should have a non-zero angular slope
proportional to the antineutrino flux.   
 A linear fit to the counting yield, $C=C_{0}-C_{1}\cos\theta$ (in the angular
 region where events from the $\nu-e$ interactions
can be neglected, see Fig. \ref{fig1}), provides the antineutrino
 flux $\Phi_{\overline{\nu}}$
 through the relation:

\begin{equation}
\Phi_{\overline{\nu}}(E_{\overline{\nu}}>E_{0})=
\frac{C_{1}}{T}\frac{2}{N_{p}\epsilon\overline{\sigma}_{0}a}
\end{equation}

\noindent
 where $N_{p}$ is the number of free protons, T is the exposure time,
 $\epsilon$
is the (assumed constant) detection efficiency, $E_{0}$ the minimal
 detectable antineutrino energy
and $\overline{\sigma}_{0}$ is the
cross section averaged over the antineutrino spectrum for
$E_{\overline{\nu}}>E_{0}$.

\par
In order to provide a quantitative illustration
of the previous points
we used data from the first 101.9 operational days of SK, as reported in 
fig.3 of \cite{tot}, corresponding to $E_{0}=8.3MeV$.
 The reported background does not show an angular
dependence. According to equation (3), to extract
 an upper limit on the solar antineutrino flux,
 we must know the average cross section
 $\overline{\sigma}_{0}$, which can be determined 
% In order to determine  $\overline{\sigma}_{0}$
% we consider
within two approches:

\smallskip
\par\noindent
$a)${\it Assuming} that the antineutrino spectrum has
 the same
 shape as that of $^{8}$B solar neutrinos, one has
 $\overline{\sigma}_{0}=7.1\cdot10^{-42}cm^{2}$. 
 This gives, as a final result,
 $\Phi_{\overline{\nu}}(E_{\overline{\nu}}>8.3MeV)<6
 \cdot 10^{4} cm^{-2}s^{-1}$,
 to the 95\% C.L.
 This bound corresponds to a fraction x=3.5\% of the solar neutrino flux
 (in the energy range $E_{\nu}>8.3MeV$) predicted by the SSM \cite{ba2}.
\smallskip
\par\noindent
$b)$ As $\sigma_{0}$ is an increasing function
of $E_{\overline{\nu}}$, one has
$\overline{\sigma}_{0}\geq\sigma_{0}(E_{0})=4.5\cdot10^{-42}cm^{2}$.
This lower limit to $\overline{\sigma}_{0}$ gives a {\it model independent}
 bound $\Phi_{\overline{\nu}}
(E_{\overline{\nu}}>8.3MeV)<9\cdot 10^{4} cm^{-2}s^{-1}$ to the 95\% C.L.

\smallskip
\par
We remark two points of the method just presented:
\begin{itemize}
\item
The sensitivity to antineutrinos increases as statistics accumulates.
Within three years
of data taking, the sensitivity
 to $\nu_{e}\rightarrow\overline{\nu}_{e}$
transition probability will reach the 1\% level, thus allowing for
a definite test of several theoretical models.
\item
The determination of the angular slope $C_{1}$
 provides a mean for {\it detecting}
antineutrinos from the Sun (and not only for deriving upper bounds).
A non vanishing slope for SK background would be, in fact, a clear
signature of a solar antineutrino flux.
\end{itemize}.

\section{Supernova detection}

 If a type--II supernova explosion occurs near the center of
our galaxy about 4000 $\overline{\nu}_{e}+p\rightarrow n+e^{+}$
events and 300 $\nu-e^{-}\rightarrow\nu-e^{-}$ events
will be produced in the Superkamiokande detector \cite{nak}.
 The detection of supernova neutrinos and antineutrinos
will provide important information on the supernova mechanism.
In particular we remark that:
    
\begin{itemize}
\item
from the strong correlation of the elastic scattering events
one can reconstruct the supernova direction with an
accuracy of about four degrees \cite{nak};
\item
from a fit to experimental data one can determine the
number of neutrinos and antineutrinos events, thus providing important inputs
to check supernova modelling.
\end{itemize}

 The anisotropy of I$\beta$D is so small that it looks
reasonable to neglect it when discussing the detector sensitivity
\cite{nak}. However, let us investigate how it can affect the
data analysis. With this aim  
 we have simulated a supernova explosion
in the galactic plane, assuming that it produces
$N_{e^+}=4000$ and $N_{e^-}=300$ events in the
detector. We took into account a $30^o$ angular resolution
\cite{nak} and repeated the simulation two-hundred times.
 We analysed the data looking for the galactic longitude
$\phi$ of the supernova and the ratio $f=N_{e^-}/N_{e^+}$
of reconstructed events, see fig.2.

\begin{figure}[htb]
\begin{center}
\vspace{-1.0cm}
\epsfig{file=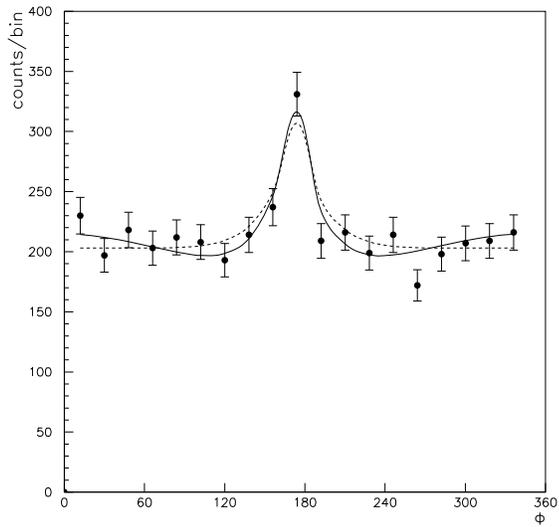,height=8cm,width=8cm}
\vspace{-1.7cm}
\end{center}
\caption[aaa]{The simulated counts as a function of the
galactic longitude for a supernova at the galactic center (dots),
with their statistical errors (bars), the best fit for $a=0$
(dotted curve) and that for $a=0.1$ (full curve)}
\label{fig2}
\end{figure}

 If the analysis is performed neglecting the anisotropy
(i.e. $a=0$) the supernova direction is correctly
determined within the predicted accuracy,
 however the reconstructed
$f$ is badly underestimated ($f=0.057\pm0.009$, 
the correct value being $f=0.075$)
and the fit is poor ($\chi^2/_{DOF}=24/18$ on the average of the 200
simulations, using 20 angular bin).
 We
 checked that when the anisotropy is included, both the
supernova direction and the ratio $f$ are correctly reconstructed,
and the quality of the fit improves significantly.
 These features are clearly understood:

\begin{itemize}

\item as the positron angular distribution is symmetrical
around the supernova direction , it does not alter its
reconstruction.

\item The depleted forward positron production mimics a
reduced number of $\nu-e$ events.

\end{itemize}

 In conclusion, we remark that inclusion of the anisotropy
of I$\beta$D is essential for extracting the correct ratio
of $\nu-e$ to $\overline{\nu_{e}}-p$ events.

%%%%%%%%%%%%%%%%%%%%%%%%%%%%%%%%%%%%%%%%%%%%%%%%%%%%%

%%%%%%%%%%%%%%%%%%%%%%%%%%%%%%%%%%%%%%%%%%%%%%%%%%%%%%
   

\begin{thebibliography}{99}
%
\bibitem{fvl}
G. Fio\-ren\-tini, M. Mo\-ret\-ti, F.L. Vil\-lan\-te, Phys. Lett. B,
 in press, 1997
%
\bibitem{tot}
Y. Tot\-su\-ka, ``First re\-sult from Su\-per-Ka\-mio\-kan\-de'',
 pre\-sen\-ted at Texas 
Sym\-po\-sium (1996).
%
\bibitem{ba2}
J.N. Bahcall and M.H. Pinsonneault, Rev. Mod. Phys. 61 (1992) 885.
\bibitem{nak}
K. Na\-ka\-mu\-ra, ''Su\-per\-ka\-mio\-kan\-de'', pre\-sen\-ted at 3rd
 NESTOR In\-ter\-na\-tio\-nal Work\-shop,
Pylos, Greece, 1993
%
\end{thebibliography}
\end{document}